\documentclass[twocolumn,
prd,aps,tightenlines]{revtex4}
\usepackage{graphicx}
\usepackage{bm}
\usepackage[]{youngtab}

\def\nn{ \nonumber \\ }

\begin{document}

\title{Baryon Exotics in the Quark Model, the Skyrme Model and QCD}

\author{Elizabeth~Jenkins}
\author{Aneesh V.~Manohar}
\affiliation{Department of Physics, University of California at San Diego,
  La Jolla, CA 92093\vspace{4pt} }

\date{\today}
\begin{abstract}
We derive the quantum numbers of baryon exotics in the Quark Model and Skyrme Model, and show that they agree for arbitrary colors and flavors. We define exoticness, $E$, which can be used to classify the states.
The exotic baryons include the recently discovered $qqqq\bar q$ pentaquarks ($E=1$), as well as exotic baryons with additional $q \bar q$ pairs ($E \ge 1$).  
The mass formula for non-exotic and exotic baryons is given as an expansion in $1/N_c$, and allows one to relate the moment of inertia of the Skyrme soliton to the mass of a constituent quark.
\end{abstract}
\maketitle

The discovery of the $\Theta^+$ baryon~\cite{Thetadiscovery} has led to renewed interest in the Quark and Skyrme Model descriptions of baryons. The collective coordinate quantization of the Skyrme Model leads to an infinite tower of states. In the three-flavor case, the lowest states of the tower for $N_c=3$ are the $(\mathbf{8},\frac 12)$ and $(\mathbf{10},\frac 32)$, which describe the non-exotic octet and decuplet baryons. The remaining states are all exotic baryons.  The first few exotic states in the baryon tower 
are the $(\mathbf{\overline {10}},\frac 1 2)$, $(\mathbf{27},\frac 12)$ and $(\mathbf{27},\frac 32)$~\cite{Manohar:1984,Chemtob}.  Since all states follow from the rotational quantization of the soliton, they have the same (positive) parity.

The $\Theta^+(1540)$ state has  quantum numbers $(\mathbf{\overline {10}},\frac 1 2)$, and the measured mass is close to the value predicted in the Skyrme Model~\cite{conf,Diakonov:1997mm,mass}. 
(Its  parity has not been measured.)  In the Skyrme Model,
the masses of the $\mathbf{8}$, $\mathbf{10}$ and $\mathbf{\overline{10}}$
are given by the soliton mass, which is the same for all multiplets in the tower, plus the rotational energy of the Skyrmion.  Thus, the mass of the $(\mathbf{\overline {10}},\frac 1 2)$ multiplet is closely related to the masses of the non-exotic octet and decuplet baryons. 

In the Quark Model, the $\Theta^+$ is an exotic baryon state which does not have the spin-flavor quantum numbers of a bound state of three constituent quarks.  The simplest baryon state with the quantum numbers of the $\Theta^+$ is a pentaquark state $uudd\bar s$~\cite{Jaffe:2003sg,quarks}. 
The mass of the $(\mathbf{\overline{10}},\frac 1 2)$  pentaquark multiplet is $5m_Q$, where $m_Q$ is the mass of a constituent quark, plus hyperfine interactions, whereas the masses of the octet and decuplet states are $3m_Q$ plus the hyperfine interactions. 

While the Skyrme and Quark Model descriptions of baryons appear to be radically different, they are, in fact, very closely related.  Both models are connected to QCD baryons by the $1/N_c$ expansion~\cite{'tHooft:1973jz}, where $N_c=3$ is the number of colors of the QCD gauge theory.
It was shown in Ref.~\cite{Manohar:1984} that all group theoretic predictions of the Quark and Skyrme Models are \emph{identical} in the $N_c \to \infty$ limit. These predictions include mass relations, ratios of axial couplings, magnetic moments, and all other quantities which do not depend on model details, such as the quark wavefunctions and the Skyrmion profile function.  The fact that the representations and group theoretic predictions of the Skyrme Model and Quark Model agree is no accident.  In Ref.~\cite{Largenspinflavor}, it was proven that
QCD baryons have an exact $SU(6)_c$ spin-flavor symmetry in the $N_c \to \infty$ limit which leads to these same group theoretic results.
The Skyrme Model and the Quark Model both contain large-$N_c$ $SU(6)_c$ symmetry,
as does QCD.  The predictions which follow from this symmetry and its breaking are model-free results which are true in QCD. 

 For finite $N_c=3$, spin-flavor symmetry breaking can be included in a systematic expansion in $1/N_c$.  The $1/N_c$ expansion of QCD gives a quantitative understanding of the spin-flavor properties of baryons~\cite{largenrefs,largen}. Results for baryon masses, magnetic moments, axial couplings, and other properties are in striking accord with experiment.  Some of the predictions are accurate at a fraction of a percent level. In addition, the $1/N_c$ expansion has been used to make very accurate predictions of $c$ and $b$ baryon masses before their discovery~\cite{EJheavybaryons}, and to understand the properties of the excited $[\mathbf{70},1^-]$ baryons~\cite{seventyminusrefs}, resolving the long-standing $\Lambda(1405)$ mass puzzle. 
By now, there is an enormous body of work which shows that $1/N_c$ can be used to make quantitative predictions in the baryon sector, and that the expansion parameter is about $1/3$, which is comparable in size to flavor $SU(3)$ breaking.

In this paper, we focus on what is common in the Quark and Skyrme Model descriptions of exotic baryons.  One of the lessons of the $1/N_c$ expansion is that the features which survive in QCD are
common to both models. Any differences between models are likely to be artifacts which are not true in QCD.  A model-free analysis of exotics directly from the $1/N_c$ expansion of QCD will be given in a subsequent publication~\cite{JM}.

We begin by studying the quantum numbers of the allowed exotic baryon states in the Skyrme and Quark Models. It turns out that the connection between the two is easiest to make if the number of light flavors is $F \ge 5$. We first present this case, and then analyze the physically relevant case $F=3$. The spin-flavor group is $SU(2F)$ under which the $F$ flavors of quarks with spin $\uparrow\downarrow$ transform as the fundamental representation. It contains the usual flavor and spin groups, $SU(2F) \supset SU(F) \otimes SU(2)$. For the realistic case of three flavors, the groups are $SU(6) \supset SU(3) \times SU(2)$.

The quantization of the Skyrme Model for arbitrary $F$ was studied in Ref.~\cite{Manohar:1984}. One can determine all the allowed representations using the methods given there~\cite{JM}.  An irreducible representation of  $SU(F)$ is described by the Dynkin weight $(n_1,n_2,\ldots,n_{F-1})$, which corresponds to a Young tableau containing $n_1$ columns of one box, $n_2$ columns with two boxes, $\ldots$, and $n_{F-1}$ columns with $[F-1]$ antisymmetrized boxes.

The only allowed representations for $F \ge 5 $ have Dynkin weights $w=(n_1,n_2,0\ldots,0,n_{-2},n_{-1})$, where we adopt the simplified notation $n_{-2} \equiv n_{F-2}$ and $n_{-1} \equiv n_{F-1}$ for convenience.   The flavor representation must satisfy
\begin{eqnarray}
\sum_k k n_k = N_c + r F, \qquad n_{-1} + n_{-2} = r ,
\label{1}
\end{eqnarray}
which defines a positive integer $r\ge0$. $N_c + r F$ is the total number of boxes. The allowed spins are $j = (n_1/2) \otimes (n_{-1}/2)$. In the Skyrme Model, each representation occurs at most once in the collective coordinate quantization. To get multiple copies of the same state, such as two $(\mathbf{8},\frac12)$ states, requires vibrational excitations of the soliton. Eq.~(\ref{1}) implies the identity
\begin{eqnarray}
n_1 + 2 n_2 -2 n_{-2} - n_{-1} &=&N_c ,
\label{2}
\end{eqnarray}
from which it follows that $j$ is integral for $N_c$ even and half-integral for $N_c$ odd. As an explicit example, for $F=8$ and $N_c=3$, the $SU(F)$ representation $(3,4,0,0,0,3,2)$ denotes the tableau
\begin{eqnarray*}
\yng(12,9,5,5,5,5,2)
\end{eqnarray*}
with $r=5$, and spins $j=3/2 \otimes 1 = 1/2 \oplus 3/2 \oplus 5/2$.

Three flavors is a special case. The Skyrme Model $SU(3)$ representations $(p,q)$ must satisfy
\begin{eqnarray}
p+2q =N_c +3 r,\qquad p+q \ge r \ge 0,
\label{4}
\end{eqnarray}
with spins 
\begin{eqnarray}
&& j = \left\{ \begin{array}{ll}{p \over 2} \otimes {r \over 2} & \text{if}\ r \le q, \\[5pt]
{q \over 2} \otimes {p+q-r \over 2} & \text{if}\  r \ge q . \\
\end{array}\right.
\label{5}
\end{eqnarray} 
For $N_c=3$, the $r=0$ states have 3 boxes and are: $(1,1)  \to (\mathbf{8},\frac12)$ and $(3,0)  \to (\mathbf{10},\frac32)$. The $r=1$ states have 6 boxes and are $(0,3) \to 
(\mathbf{10},\frac12)$, $(2,2) \to (\mathbf{27}, \frac 12),(\mathbf{27}, \frac 32)$, $(4,1) \to (\mathbf{35},\frac32),(\mathbf{35},\frac52)$, $(6,0) \to (\mathbf{28},\frac52) $. In Ref.~\cite{Diakonov:2003ei}, $r$ was called exoticness, $E$. We shall see below that this is not the appropriate definition of $E$, and that the $ (\mathbf{28},\frac52)$ is an $E=2$ state even though it has $r=1$.

Quark Model exotic baryons have $N_c+E$ quarks and $E$ antiquarks, where exoticness $E$ is the  number of $q \bar q$ pairs of the leading Fock component of the state.  Pentaquarks are $E=1$ states, and non-exotic states are included as the special case $E=0$.

A dominant feature of the $1/N_c$ analysis of baryons is the importance of spin-flavor quantum numbers. Spin-flavor interactions are generated by pion exchange, which is explicit in the Chiral Quark Model~\cite{Manohar:1983md}.  In contrast, gluon exchange generates spin-color interactions. Although it has been argued that spin-flavor interactions are the dominant interactions between quarks, it is not easy to distinguish the two. In the ground state baryons, spin-flavor or spin-color forces lead to the same hyperfine splittings for the masses. However, one can distinguish between the two by looking at the $[\mathbf{70},1^-]$, where there are some indications that spin-flavor interactions dominate, though the question is far from settled~\cite{seventyminusrefs}.  We will assume that the the $N_c+E$ quarks (and the $E$ antiquarks) are in completely symmetric $SU(2F)$ spin-flavor states, as discussed in previous work~\cite{Glozmanetc}, and so are completely antisymmetric in color-orbital space. In the Quark Model, this implies that $E$ quarks are in an orbitally excited state with $l=1$, so that the baryon still has positive parity, as in the Skyrme Model.

\begin{figure}
\begin{center}
\includegraphics[width=3cm]{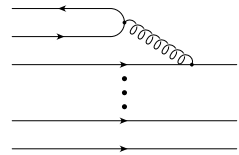}
\end{center}
\caption{Annihilation of a $q \bar q$ pair in a baryon.\label{fig1}}
\end{figure}

In large-$N_c$ QCD, there is a $U(2F)_q \times U(2F)_{\bar q}$ spin-flavor symmetry which acts separately on the quarks and antiquarks~\cite{JM}. This symmetry is broken to the diagonal $U(F)\times SU(2)$ subgroup by quark-antiquark pair creation/annihilation at order $1/\sqrt{N_c}$ (see Fig.~\ref{fig1}). The suppression factor is only $\sqrt{N_c}$, since the graph contains half an additional closed quark loop.
A flavor singlet $\bar q q$ pair can be either $J^P =0^-$ or $1^-$.
In the meson sector, annihilation effects (which are of order $1/N_c$ rather than $1/\sqrt{N_c}$) are small in the vector channel, where the states form ideally mixed nonets, but are much larger in the pseudoscalar channel, as evidenced by the large mass splitting of the octet $\eta$ and singlet $\eta^\prime$.  [Instanton effects contribute in the pseudoscalar channel.]   In the baryon sector, both the $0^-$
and $1^-$ channels for flavor singlet $\bar q q$ pair annihilation will contribute significantly, since a $1^-$ $\bar q q$ pair can transfer its angular momentum to the rest of the baryon and then annihilate in the $0^-$ channel. Thus, we expect that flavor singlet $q \bar q$ annihilation is an important dynamical effect for exotic baryons.  The fact that the presently discovered $\Theta^+(1540)$ and $\Xi^{--}_{\frac 3 2}(1860)$~\cite{xiexotics} of the antidecuplet are the ones which cannot decay by flavor singlet $q \bar q$ annihilation may be a manifestation of this strong dynamics.

The completely symmetric $(N_c+E)$ quark state transforms as the 
\Yvcentermath1
\begin{eqnarray}
(N_c+E,0,\ldots,0)=\yng(2)\cdots\yng(2)\quad (N_c+E\ \text{boxes})
\label{6}
\end{eqnarray}
representation of $SU(2F)_q$, and the $E$ antiquark state transforms as the 
\begin{eqnarray}
(0,\ldots,0,E)= \overline{\yng(2)\cdots\yng(2)}\quad (E\ \text{boxes})
\label{7}
\end{eqnarray}
representation of $SU(2F)_{\bar q}$. Under $SU(2)\times SU(F)$, the quark spin-flavor representation breaks up into
\begin{eqnarray}
j=\frac12&& \yng(1,1)\cdots\yng(4,3) \nn
j=\frac32&&\yng(1,1)\cdots\yng(5,2) \nn
j=\frac52&& \yng(1,1)\cdots\yng(6,1) \nn
&& \quad\qquad \vdots \nn
j=\frac{N_c+E}2&& \yng(7) 
\label{8}
\end{eqnarray}
and similarly for the antiquark representation. Thus quarks and antiquarks give the states $(n_1,n_2,0,\ldots,0)$ and $(0,\ldots,0,n_{-2},n_{-1})$ of $SU(F)$, respectively, with
\begin{eqnarray}
n_1+2 n_{-1} = N_c + E, \qquad n_{-1} + 2 n_{-2} = E
\label{9}
\end{eqnarray}
and spins 
\begin{eqnarray}
j_q=n_1/2,\qquad j_{\bar q}=n_{-1}/2.
\label{10}
\end{eqnarray}

The exotic baryon states of the Quark Model are given by combining the quarks with the antiquarks. There are many possible flavor representations which result. Again, states in which a quark can annihilate with an antiquark are dropped, which
corresponds to using the minimal quark content for a given flavor representation, something that is routinely done in most analyses. For example, the proton is treated as a $uud$ state, rather than as a superposition of states $uud (\bar q q)^E$ with $E\ge 0$. One cannot distinguish a flavor singlet $q \bar q$ pair from gluons or orbital angular momentum.
 In the $1/N_c$ analysis, one does not have to make any assumption about the quark content of the baryon, since the higher Fock components are automatically included via $1/N_c$ corrections to the coefficients of lower-body operators. The exoticness $E$ of a baryon is defined as the \emph{minimum} number of $q \bar q$ pairs needed to construct a baryon state with the specified quantum numbers.

Combining the quark and antiquark states in Eqs.~(\ref{6},\ref{7}) with the rule that flavor representations which can annihilate into lower $E$ states are dropped (i.e. there are no flavor index contractions) gives precisely the same states as in the Skyrme Model, Eq.~(\ref{1}). The Quark Model gives a physical  interpretation of the the integers $n_i$ of the exotic baryon flavor representation in terms of $j_q$, $j_{\bar q}$, $N_c$ and $E$ using Eqs.~(\ref{9},\ref{10}).
Although the present derivation applies for $F \ge 5$, 
the representations also match for  three flavors, as will be discussed at the end of this article. 

Thus, our first result is that the allowed baryon states in the Skyrme and Quark Models are identical. The Skyrme Model naturally knows about $N_c$ and $r$ through Eq.~(\ref{1}). The Quark Model naturally knows about $N_c$ and $E$ via Eq.~(\ref{9}).
Note that $E \ge r$, and $E-r=n_{-2}$. Naively, one might view the extra $rF$ boxes in the Skyrme Model as representing $r$ $\bar qq$ pairs, since the quark representation $\yng(1)$ has a single box, and the antiquark representation $\overline{\yng(1)}$
has $F-1$ boxes. However, this is not correct; the number of $q \bar q$ pairs is given by $E$. To understand this distinction, consider the tensor product of two antiquark flavor representations
\begin{eqnarray*}
\overline{\yng(1)} \otimes \overline{\yng(1)} = \overline{\yng(2)} + \overline{\yng(1,1)} \ .
\end{eqnarray*}
The left-hand side has $2(F-1)$ boxes, as does the first tensor product representation.  The second representation on the right-hand side has only $F-2$ boxes, however, because a $[F]$ column, which is a singlet, has been dropped.  For the general Dynkin weight
$w$, $n_{-2}$ $[F]$ columns have been dropped, so that $E=r+n_{-2}$, which gives the relation for $E$ in Eq.~(\ref{9}), using Eq.~(\ref{1}). 

In the Quark Model, the main contribution to the mass of an exotic baryon is $(N_c+2E) m_Q$, where $m_Q$ is the constituent quark mass. To this mass, one adds a contribution from the hyperfine interactions among quarks, proportional to ${\mathbf J}_q^2=j_q(j_q+1)$, and among antiquarks,  proportional to ${\mathbf J}_{\bar q}^2=j_{\bar q}(j_{\bar q}+1)$, and between quarks and antiquarks proportional to $\left(\mathbf{J}_q \cdot \mathbf{J}_{\bar q}\right)
=(\mathbf{J}^2-\mathbf{J}_q^2-\mathbf{J}_{\bar q}^2)/2$, where $\mathbf{J}^2=j(j+1)$ and $j$ is the total spin of the state. We now show that the Skyrme Model rotational energy reproduces the Quark Model form for the energy, even though, as pointed out earlier, $E$ is not a natural variable for the Skyrme Model.

The Hamiltonian of the Skyrme Model is
\begin{eqnarray}
H = M_0 +{1 \over 2 I_1}\mathbf{J}^2 +  {1 \over 2 I_2} \left( 
\mathbf{T}^2 - \mathbf{J}^2 - {F-2 \over 4 F} N_c^2\right),
\label{13}
\end{eqnarray}
where $M_0$, the soliton mass, and $I_{1,2}$, the moments of inertia, are of order $N_c$. The Casimir $\mathbf{T}^2$ of the flavor representation $(n_1,n_2,0,\ldots,0,n_{-2},n_{-1})$ can be computed explicitly, and the result is not very illuminating. However, if we change variables from $n_i$ to $j_q$, $j_{\bar q}$, $N_c$ and $E$ (\emph{rather than $r$}), one finds
\begin{eqnarray}
{\mathbf  T}^2 \! &=&\!  \frac 1 2 E(E + 2 F+N_c -4)+ {N_c( N_c+2F)(F-2) \over 4 F}
+ \mathbf{J}^2_q + \mathbf{J}^2_{\bar q} \cr
H &=& M_0 + {1 \over 2 I_1}\mathbf{J}^2
+ {1 \over 2 I_2}\times \nn
&&\hspace{-1cm} \biggl[ \frac 1 2 E(E + 2 F+N_c -4)
- \frac 1 2 (F-2) N_c -2 \left( \mathbf{J}_q \cdot  \mathbf{J}_{\bar q} \right) \biggr] .
\label{14}
\end{eqnarray}
Eq.~(\ref{14}) has two important features: (1) the order $N_c^0$ term $E N_c/(4I_2)$ is linear in $E$; (2) the terms proportional to $E^2$, $\mathbf{J}^2$ and $\mathbf{J}_q \cdot \mathbf{J}_{\bar q}$ are order $1/N_c$.  These features also follow from a general $1/N_c$ analysis.
The variable $E$ in Eq.~(\ref{14}) is the number of $\bar qq$ pairs.  In the Quark Model, each $q\bar q$ pair adds $2 m_Q$  to the energy, where $m_Q$ is of order $N_c^0$.   The relation between the constituent quark mass of the Quark Model and the moment of inertia of the Skyrme Model implied by Eq.~(\ref{14}) is
\begin{eqnarray}
m_Q &=& {N_c \over 8 I_2},
\label{15}
\end{eqnarray}
which is a truly remarkable result---the Skyrme Model knows about constituent quarks!  Eq.~(\ref{14}) was used to predict the $\Theta^+$ mass to be around 
1540~MeV in Ref.~\cite{Diakonov:1997mm}; using  their value $I_2 = (500\  \text{MeV})^{-1}$  gives $200$~MeV for the constituent quark mass. If instead we define $2 m_Q$ using the entire term linear in $E$ in Eq.~(\ref{14}), then $N_c \to N_c+2F-4$ in Eq.~(\ref{15}), and $m_Q=300$~MeV for $F=3$.  In some recent papers, Eq.~(\ref{14}) for exotics has been shown to be invalid~\cite{Itzhaki:2003nr,Cohen:2003mc}. Nevertheless, the form Eq.~(\ref{14}) is still valid, though the coefficients of the individual terms, which can be computed in a systematic semiclassical expansion in the Skyrme Model, are not given simply by $I_1$ and $I_2$ because of vibrational-rotation coupling. Computing the term linear in $E$ will still give a constituent quark mass of order $N_c^0$, but the simple relation Eq.~(\ref{15}) has corrections of order one~\footnote{Note that terms linear in time derivatives, which are proportional to $\epsilon_{\mu\nu\alpha\beta}$, break the $F(r) \to -F(r)$ symmetry of the interaction between Skyrmions and heavy mesons. As a result Skyrmion-$D,\bar B$ interactions differ from Skyrmion-$\bar D,B$ interactions, and it is not possible to relate the mass of heavy exotics such as $\Theta_{c,b}$ to the $\Lambda_{c,b}$~\cite{boundstate}.}.

The above analysis was performed for $F \ge 5$.  For $F=3$, $n_{-2}$ and $n_1$ both count columns with one box, and $n_{-1}$ and $n_2$ both count columns with two boxes. [This is just the statement that the antisymmetric product  $(\mathbf{3} \times \mathbf{3})_A=\overline{\mathbf{3}}$.] As a result, the quark and antiquark contributions to the $SU(3)$ weight $(p,q)$ are not nicely separated, as they are for $F \ge 5$. The Skyrme Model states for $F=3$ are listed in Eqs.~(\ref{4},\ref{5}). In the Quark Model, the $N_c+E$ quarks give states $(n_1,n_2)$ with spin $j_q=n_1/2$; the $E$ antiquarks gives states $(n_{-2},n_{-1})$ with spin $j_{\bar q} = n_{-1}/2$, and the $n_i$ satisfy Eq.~(\ref{9}). The baryon representation $(p,q)$ for $F=3$ is 
\begin{eqnarray}
p=n_1 + n_{-2},&\qquad& q=n_2+n_{-1},
\label{16}
\end{eqnarray}
where $p + 2 q \equiv N_c + 3r$, so that $r=n_{-1}+n_{-2}$ and Eq.~(\ref{4}) is satisfied.  The allowed
spins are $j=(n_1)/2 \otimes (n_{-1})/2$, but now knowing $(p,q)$ does not determine all four $n_i$ uniquely, as it did for $F\ge5$. However, we can use the principle of minimal flavor content---that one chooses the minimum value of exoticness $E$ which produces flavor state $(p,q)$.  The solution of Eqs.~(\ref{16}) which satisfies Eq.~(\ref{9}) and minimizes $E$ is
\begin{eqnarray}
&& (n_1,n_2,n_{-1},n_{-2},E)\nonumber\\[5pt]
& = &  \left\{ \begin{array}{ll} (p,q-r,r,0,r) & \text{if}\ r \le q, \\
(p+q-r,0,q,r-q,2r-q) & \text{if}\ r \ge q .\\
\end{array}\right.
\label{17}
\end{eqnarray}
With this choice, the spins $j=j_q \otimes j_{\bar q}$ are the same as Eq.~(\ref{5}), so the Quark and Skyrme Model states are the  \emph{same}, and Eq.~(\ref{14}) still holds.
Eq.~(\ref{17}) allows one to determine the exoticness $E$, as well as the quark and antiquark content of a given exotic baryon from its $SU(3)$ representation $(p,q)$. 
The $\mathbf{8},\mathbf{10}$ are $E=0$; $\mathbf{\overline{10}},\mathbf{27}, \mathbf{35}$ are $E=1$; and $\mathbf{28}$ is $E=2$,  even though it corresponds to an $r=1$ tableau with 6 boxes. One cannot construct an $SU(3)$ $\mathbf{28}$ from $q^4 \bar q$.

The analysis of Ref.~\cite{Jaffe:2003sg} gave a degenerate $(\mathbf{8},\frac12)$ and $(\mathbf{\overline{10}},\frac12)$ for $q^4 \bar q$. We would argue that the $q^4\bar q$ $(\mathbf{8},\frac12)$, which has $E=0$ using Eq.~(\ref{17}), can annihilate into the normal $qqq$ $(\mathbf{8},\frac12)$ baryons, and that this is  a significant effect. When $SU(3)$ breaking is included, the 7 states in the $(\mathbf{\overline{10}},\frac12)$ which are not at the corners of the triangle can mix with the corresponding states in the $qqq$
$(\mathbf{8},\frac12)$ at order $m_s/\sqrt{N_c}$, raising their energy, and changing the simple picture of an $SU(3)$ $\mathbf{\overline{10}}$ of exotics. It is important to look for these 7 states as they will provide information about the size of the $m_s/\sqrt{N_c}$ annihiliation mixing. This information can then be used to estimate the mixing between the $qqq$ and $q^4\bar q$ $(\mathbf{8},\frac12)$ baryons, which is not suppressed by $m_s$.

The $\Delta \to N \pi$ coupling constant $g_A$ is order $N_c$, whereas the $\Theta^+ \to NK $ coupling $g_{\Theta N}$ is order $\sqrt{N_c}$~\cite{largenrefs}. In the large $N_c$ limit, the $\Delta$--$N$ mass difference is order $1/N_c$, so the $p^3$ phase-space makes the $\Delta \to N\pi$ width of order $g_A^2 p^3/f_\pi^2 \sim 1/N_c^2$~\cite{largenrefs}. The $\Theta$--$N$ mass difference is order one, so the $\Theta^+ \to NK $ width is of order $g_{\Theta N}^2 p^3/f_\pi^2 \sim N_c^0$~\cite{Praszalowicz:2003tc}. Nevertheless, the width can be narrow. In the $1/N_c$ expansion, the $N$ and $\Theta$ are members of two different irreducible representations, and the $\Theta^+ \to NK $ coupling should be compared with the $N^* N \pi$ coupling, which is much smaller than the $NN \pi$ coupling. This additional suppression can easily give a width below 10~MeV, but relies on a model estimate of the unknown coupling.

\acknowledgments

This work was supported in part by the Department of Energy under Grant 
DE-FG03-97ER40546.

\end{document}